# A diquark model for baryons containing one heavy quark

D. Ebert*,[1], T. Feldmann*,**,[1], C. Kettner***,[2], and H. Reinhardt***,[2]

[1] Institut für Physik, Humboldt-Universität zu Berlin, Invalidenstraße 110, D-10115 Berlin, Germany
[2] Institut für Theoretische Physik, Universität Tübingen, Auf der Morgenstelle 14, D-72076 Tübingen, Germany

June 1995

**Abstract.** We present a phenomenological ansatz for coupling a heavy quark with two light quarks to form a heavy baryon. The heavy quark is treated in the heavy mass limit, and the light quark dynamics is approximated by propagating scalar and axial vector 'diquarks'. The resulting effective lagrangian, which incorporates heavy quark and chiral symmetry, describes interactions of heavy baryons with Goldstone bosons in the low energy region. As an application, the Isgur-Wise form factors are estimated.

## 1 Introduction

The physics of particles built up with one heavy quark becomes much simpler in the heavy quark limit due to the subsequent heavy flavor and spin symmetry (see e.g. [1] and references therein). These symmetries are realized as long as the heavy quark mass $m_Q$ is much larger than a typical QCD scale, $m_Q \gg \Lambda_{QCD}$, which is the case for $Q = b, c$. Phenomenologically, they show up as an approximate degeneracy of heavy hadrons differing only in the spin of the heavy quark, and as a definite scaling of observables with the heavy quark masses. Since the heavy quark in the infinite mass limit is to be viewed as a static source of color moving with some fixed velocity $v_\mu$, it is the remaining light degrees of freedom which determine the hadronic properties. In Table 1 we present the heavy baryon mass spectrum (using some estimates as given for example in [2]) labeled by isospin $I$ and strangeness $S$[1].

Although Heavy Quark Effective Theory (HQET) provides the tool for calculating systematically the perturbative QCD corrections, one is left with the non-perturbative features of QCD that lead to the formation of hadrons (containing light quarks) at low energies. The low energy properties of light quark systems on the other hand are dictated by the approximate chiral symmetry of QCD and its spontaneous breaking, explaining the emergence of very light pseudoscalar mesons as Goldstone particles in the physical spectrum. The success of this is due to the fact that the light quark masses ($q = u, d, s$) are small compared to the QCD scale, $m_q \ll \Lambda_{QCD}$.

The interplay between chiral and heavy quark symmetry is thus studied within effective lagrangians for strong and weak interactions of hadrons with the various contributing terms restricted by symmetries [3, 4, 5, 6]. The coefficients in front of these terms are in principal free parameters to be fixed by experiment, but have also been related to microscopic quark models for light mesons [7] as well as for heavy mesons (containing one heavy quark) [8, 9, 10]. Light baryons have been recently described within quark models as bound states of quarks and diquarks, resulting in Faddeev type of equations [11, 12, 13]. This paper is a first step to obtain an analogous description of heavy baryons (containing one heavy quark) in the heavy quark limit by making use of the diquark picture. Light diquarks are coupled to the heavy quark by a simple model interaction.

The effective lagrangian for strong interactions of heavy baryons can then be obtained directly by functional methods. The introduction of weak heavy quark currents also offers the possibility of determining the Isgur-Wise functions which are the universal form factors for heavy quark transitions inside the heavy baryon (for a general discussion see for example [14]). Weak form factors and strong coupling constants can then be used to describe strong and weak semileptonic decays of heavy baryons like $\Sigma_Q \to \Lambda_Q + \pi$, $\Lambda_b \to \Lambda_c + l\nu$, $\Omega_b \to \Omega_c + l\nu$, $\Sigma_b \to \Lambda_c + \pi + l\nu$.

The organization of this paper is as follows: In the next section we discuss some general properties of heavy baryons related to heavy quark and chiral symmetry. In section 3 we will define our model from which the effective heavy baryon lagrangian is derived in section 4. Relations for heavy baryon masses, coupling constants with light mesons and Isgur-Wise form factors are discussed in some detail.

* Supported by *Deutsche Forschungsgemeinschaft* under contract Eb 139/1-1.
** E-mail: FELDMANN@PHA2.PHYSIK.HU-BERLIN.DE
*** Supported by COSY under contract 41170833.

[1] The notation is explained later on in the text.

Table 1. Heavy baryons: masses ([MeV] in parentheses) and flavor content. The mass differences between the first two columns measure the deviation from the spin symmetry limit. In the flavor symmetry limit the bottom baryon masses should differ from those of the charmed ones only by a constant equal to $m_b - m_c$.

| charmed baryon | spin partner | bottom baryon | notation | $ij$ | I | S |
| --- | --- | --- | --- | --- | --- | --- |
| $\Lambda_c^+$ (2285) | trivial | $\Lambda_b^0$ (5641) | $T_v^{ij}$ | $[ud]$ | 0 | 0 |
| $\Xi_c^{+,0}$ (2470) | trivial | $\Xi_b^{0,-}$ ($\approx 5800$) | $T_v^{ij}$ | $[us]$, $[ds]$ | 1/2 | -1 |
| $\Sigma_c^{++,+,0}$ (2453) | $\Sigma_c^{*++,+,0}$ (2530) | $\Sigma_b^{(*)+,0,-}$ ($\approx 5820$) | $S_v^{\mu\,ij}$ | $uu$, $\{ud\}$, $dd$ | 1 | 0 |
| $\Xi_c^{\prime+,0}$ ($\approx 2570$) | $\Xi_c^{*+,0}$ ($\approx 2630$) | $\Xi_b^{\prime(*)0,-}$ ($\approx 5940$) | $S_v^{\mu\,ij}$ | $\{su\}$, $\{ds\}$ | 1/2 | -1 |
| $\Omega_c^0$ (2719) | $\Omega_c^{*0}$ ($\approx 2740$) | $\Omega_b^{(*)-}$ ($\approx 6050$) | $S_v^{\mu\,ij}$ | $ss$ | 0 | -2 |

## 2 Chiral and heavy quark symmetry for heavy baryons

In the heavy quark limit it is sufficient to deal with the heavy quark spinors projected on the heavy quark velocity $v_\mu$

$$Q_v(x) = \frac{1+\slashed{v}}{2} e^{im_Q v \cdot x} Q(x) \,, \quad v^2 = 1 \,, \tag{1}$$

where the heavy mass dependence has been made explicit. The heavy quark part of the QCD Lagrangian then simplifies to

$$\mathcal{L}_{\text{HQET}} = \bar{Q}_v (iv \cdot \mathcal{D}[G_\mu]) Q_v + O(1/m_Q) \tag{2}$$

where $\mathcal{D}[G_\mu]$ is the covariant derivative with respect to the gluon field. As a consequence, physics of strong interactions in the heavy quark limit becomes independent of the heavy quark spin and flavor and conserves the heavy quark velocity. Therefore the two light quarks determine the relevant quantum numbers of the heavy baryon. The coupling of the color, flavor[2] and spin degrees of freedom of the two light quarks forming the diquark reads

$$3_C \otimes 3_C = \bar{3}_C \oplus 6_C \,, \tag{3}$$
$$3_F \otimes 3_F = \bar{3}_F \oplus 6_F \,, \tag{4}$$
$$(1/2) \otimes (1/2) = 0 \oplus 1 \,. \tag{5}$$

In the following, we only consider the color anti-triplet part in (3) which together with the color triplet of the heavy quark forms the physical color singlet baryon state. The Pauli principle then requires the diquark fields to be symmetric with respect to simultaneous exchange of spin and flavor indices. Therefore scalar ($0^+$) diquarks only occur in the flavor anti-triplet $\bar{3}_F$ and are hence realized by an antisymmetric flavor matrix

$$D^{ij} = \frac{i}{\sqrt{2}} \begin{pmatrix} 0 & D^{[ud]} & D^{[us]} \\ -D^{[ud]} & 0 & D^{[ds]} \\ -D^{[us]} & -D^{[ds]} & 0 \end{pmatrix} \,. \tag{6}$$

Analogously, axial-vector ($1^+$) diquarks are realized by the symmetric $6_F$ flavor matrix

$$F^{ij} = \frac{1}{\sqrt{2}} \begin{pmatrix} \sqrt{2} F^{uu} & F^{\{ud\}} & F^{\{us\}} \\ F^{\{ud\}} & \sqrt{2} F^{dd} & F^{\{ds\}} \\ F^{\{us\}} & F^{\{ds\}} & \sqrt{2} F^{ss} \end{pmatrix} \,. \tag{7}$$

To include the interactions with light mesons it is convenient to implement chiral symmetry in the usual nonlinear realization $\xi(x) = \exp(i\pi(x)/F_\pi)$ where $F_\pi \approx$ 93 MeV is the weak meson decay constant, and $\pi(x) = \pi^a(x)\lambda^a/2$ denotes the light pseudoscalar meson octet with $\lambda^a$ ($a = 1\ldots 8$) being the Gell–Mann matrices, tr$[\lambda^a \lambda^b] = 2\delta^{ab}$. Under global $SU(3)_L \otimes SU(3)_R$ chiral transformations, these fields transform as

$$\xi \to L\xi U^\dagger = U\xi R^\dagger \tag{8}$$

which defines the chiral matrix $U = U(L, R, \xi(x))$ as a non-linear function of $\xi$ and represents local transformations in $SU(3)_V$. Matter fields then couple to the vector and axial vector combinations

$$\mathcal{V}_\mu = \frac{i}{2}(\xi^\dagger \partial_\mu \xi + \xi \partial_\mu \xi^\dagger) \to U\mathcal{V}_\mu U^\dagger + iU\partial_\mu U^\dagger \,, \tag{9}$$

$$\mathcal{A}_\mu = \frac{i}{2}(\xi^\dagger \partial_\mu \xi - \xi \partial_\mu \xi^\dagger) \to U\mathcal{A}_\mu U^\dagger \,. \tag{10}$$

The vector field $\mathcal{V}_\mu$, transforming as a gauge field of $SU(3)_V$, couples via a properly defined covariant derivative[3]. The coupling constants to the axial field $\mathcal{A}_\mu$ are not restricted by chiral symmetry.

From this it follows that the diquark fields transform as

$$D^{ij} \to U^{ii'} D^{i'j'} U^{jj'} = (UDU^T)^{ij} \,, \tag{11}$$

$$F^{\mu ij} \to U^{ii'} F^{\mu i'j'} U^{jj'} = (UF^\mu U^T)^{ij} \,, \tag{12}$$

and their covariant derivative in $SU(3)_V$ is defined as

$$\mathcal{D}_\mu D = \partial_\mu D - i\mathcal{V}_\mu D - iD\mathcal{V}_\mu^T \,, \tag{13}$$

$$\mathcal{D}_\nu F^\mu = \partial_\nu F^\mu - i\mathcal{V}_\nu F^\mu - iF^\mu \mathcal{V}_\nu^T \,,$$
$$F^{\mu\nu} = \mathcal{D}^\mu F^\nu - \mathcal{D}^\nu F^\mu \,. \tag{14}$$

Consequently, in the heavy mass limit we have two types of heavy baryons:

i) The first one contains a heavy quark together with the two light quarks transforming as the scalar diquark $D^{ij}$. We therefore obtain Dirac spinors for spin-1/2 particles,

$$T_v^{ij} = \frac{i}{\sqrt{2}} \begin{pmatrix} 0 & \Lambda_Q & \Xi_{Q, I_3=\frac{1}{2}} \\ -\Lambda_Q & 0 & \Xi_{Q, I_3=-\frac{1}{2}} \\ -\Xi_{Q, I_3=\frac{1}{2}} & -\Xi_{Q, I_3=-\frac{1}{2}} & 0 \end{pmatrix} \tag{15}$$

where $v$ denotes the projection on the heavy quark velocity $\slashed{v} T_v = T_v$. The two spin orientations of $T_v$ form the (in this case trivial) spin symmetry partners.

---

[2] Throughout we consider the flavor group $SU(3)_F$.

[3] The light vector meson octet can be introduced analogously by the concept of hidden symmetry [15].

ii) Coupling the heavy quarks with the light quarks transforming as the axial vector diquark $F^{\mu\,ij}$, one gets either spin-1/2 Dirac fields $B_v^{ij}$ or spin-3/2 Rarita–Schwinger fields $B_v^{*\,\mu\,ij}$. These are then symmetry partners with respect to spin rotations of the heavy quark. Consequently, they can be combined in a multiplet

$$S_v^{\mu\,ij} = \frac{1}{\sqrt{3}}\,\gamma_5(\gamma^\mu - v^\mu)\,B_v^{ij} + B_v^{*\,\mu\,ij}\,, \qquad (16)$$

$$B_v^{(*)\,ij} =$$
$$\frac{1}{\sqrt{2}}\begin{pmatrix} \sqrt{2}\Sigma_{Q,I_3=1}^{(*)} & \Sigma_{Q,I_3=0}^{(*)} & \Xi'^{(*)}_{Q,I_3=\frac{1}{2}} \\ \Sigma_{Q,I_3=0}^{(*)} & \sqrt{2}\Sigma_{Q,I_3=-1}^{(*)} & \Xi'^{(*)}_{Q,I_3=-\frac{1}{2}} \\ \Xi'^{(*)}_{Q,I_3=\frac{1}{2}} & \Xi'^{(*)}_{Q,I_3=-\frac{1}{2}} & \sqrt{2}\Omega_Q^{(*)} \end{pmatrix} \qquad (17)$$

with

$$v_\mu S_v^\mu = 0\,, \quad \slashed{v} S_v^\mu = S_v^\mu\,. \qquad (18)$$

Heavy flavor symmetry further relates the residual masses $\Delta M_T = M_T - m_Q$ for different heavy flavors $Q$, and similiarly for $\Delta M_S = M_S - m_Q$. The transformation of the heavy baryon fields and their covariant derivatives under chiral symmetry is given by the light quark content,

$$T_v \to U T_v U^T\,,$$
$$\mathcal{D}_\mu T_v = \partial_\mu T_v - i\mathcal{V}_\mu T_v - iT_v \mathcal{V}_\mu^T \to U \mathcal{D}_\mu T_v U^T\,, \quad (19)$$
$$S_v^\mu \to U S_v^\mu U^T\,,$$
$$\mathcal{D}_\nu S_v^\mu = \partial_\nu S_v^\mu - i\mathcal{V}_\nu S_v^\mu - iS_v^\mu \mathcal{V}_\nu^T \to U \mathcal{D}_\nu S_v^\mu U^T\,. (20)$$

## 3 A model with heavy quarks and light diquarks

The model we are proposing treats the two light quarks within the heavy baryons as effectively propagating scalar and axial vector pointlike diquarks. This is sufficient to obtain a dynamics that respects chiral and heavy quark symmetry. The model lagrangian reads

$$\mathcal{L} = \mathcal{L}_0 + \mathcal{L}_1 + \mathcal{L}_2\,, \qquad (21)$$
$$\mathcal{L}_0 = \bar{Q}_v iv\cdot\partial Q_v$$
$$\quad + \mathrm{tr}\,[(\mathcal{D}_\mu D^\dagger)(\mathcal{D}^\mu D) - M_D^2 D^\dagger D]$$
$$\quad - \frac{1}{4}\mathrm{tr}\,[F_{\alpha\beta}^\dagger F^{\alpha\beta}] + \mathrm{tr}\,[M_F^2 F_\mu^\dagger F^\mu]\,, \qquad (22)$$
$$\mathcal{L}_1 = \lambda_1\sqrt{M_D M_F}\,\mathrm{tr}\,[D^\dagger \mathcal{A}_\nu F^\nu + h.c.]$$
$$\quad + \lambda_2\,\mathrm{tr}\left[\frac{1}{2}\epsilon^{\mu\sigma\alpha\beta}F_\mu^\dagger \mathcal{A}_\sigma F_{\alpha\beta} + h.c\right]\,, \qquad (23)$$
$$\mathcal{L}_2 = \tilde{G}_1\,\mathrm{tr}\,[\bar{Q}_v D^\dagger D Q_v]$$
$$\quad - \tilde{G}_2\,\mathrm{tr}\,[\bar{Q}_v F_\mu^\dagger(g^{\mu\nu} - v^\mu v^\nu)F_\nu Q_v]\,, \qquad (24)$$

where we omitted flavor and color indices for transparency. Here $\mathcal{L}_0$ includes the kinetic terms of the heavy quark and the scalar ($D$) and axial vector ($F^\mu$) diquarks with $M_{D,F}$ being their effective masses. In $\mathcal{L}_1$ their axial charges $\lambda_{1,2}$ enter as effective parameters to describe the interaction with soft pions, kaons and $\eta$-mesons. Note that the process $D \to D + \pi$ violates parity and consequently the amplitude for $T_v \to T_v + \pi$ vanishes in the heavy quark limit.

Finally, $\mathcal{L}_2$ contains the model ansatz for a local coupling of light diquarks and heavy quarks. The transversality condition (18) has been taken into account explicitly in the interaction of the heavy quark with the axial vector diquark by introducing a transversal projector $P_{\mu\nu}^\perp = g_{\mu\nu} - v_\mu v_\nu$. The effective couplings $\tilde{G}_1$ and $\tilde{G}_2$ have dimension mass$^{-1}$: $\tilde{G}_1 = \tilde{g}_1/\Lambda$, $\tilde{G}_2 = \tilde{g}_2/\Lambda$ introducing an inherent cut-off scale $\Lambda$ which we suppose to be of the order of 1 GeV. Note that because of the field transformation (1) this cut-off effects the residual momentum of the heavy quark $k^\mu = p^\mu - m_Q v^\mu$ which is of the order of the light quark momenta.

For an explicitly broken $SU(3)_F$, $m_s \neq m_u \approx m_d$, the parameters in (21) depend on the various channels characterized by isospin $I$ and strangeness $S$ in Table 1.

The lagrangian can as usual be rewritten by introducing collective baryonic fields $T_v(x)$, $S_v^\mu(x)$

$$\mathcal{L}_2 \to \mathrm{tr}\,[\bar{Q}_v D^\dagger T_v + h.c.] - \frac{1}{\tilde{G}_1}\mathrm{tr}\,[\bar{T}_v T_v]$$
$$\quad + \mathrm{tr}\,[\bar{Q}_v F_\mu^\dagger S_v^\mu + h.c.] + \frac{1}{\tilde{G}_2}\mathrm{tr}\,[\bar{S}_{v\,\mu} S_v^\mu]\,, \qquad (25)$$

where the spin-1/2 and spin-3/2 part of the multiplet field $S^\mu$ can be identified by using the completeness relation in the heavy mass limit

$$(g_{\mu\nu} - v_\mu v_\nu) = P_{\mu\nu}^{3/2} + P_{\mu\nu}^{1/2}\,, \qquad (26)$$
$$P_{\mu\nu}^{1/2} = \frac{1}{3}(\gamma_\mu - \slashed{v}v_\mu)(\gamma_\nu - \slashed{v}v_\nu)\,, \quad v^\mu P_{\mu\nu}^{1/2} = 0\,, \qquad (27)$$
$$v^\mu P_{\mu\nu}^{3/2} = \gamma^\mu P_{\mu\nu}^{3/2} = 0\,, \qquad (28)$$

where $P_{\mu\nu}^{1/2}$ projects out the transversal spin-1/2 part and $P_{\mu\nu}^{3/2}$ the spin-3/2 part.

After the transformation (25) it is straightforward to integrate out first the heavy quark field and subsequently the scalar and axial-vector diquarks. This leads to

$$\det[\mathcal{M}]^{-1}\,, \qquad (29)$$

with

$$\mathcal{M}_{11} = \Delta_D^{-1} + (iv\cdot\partial)^{-1}\,T_v \bar{T}_v\,,$$
$$\mathcal{M}_{12} = \sqrt{M_D M_F}\,\lambda_1\,\mathcal{A}^\nu + (iv\cdot\partial)^{-1}\,T_v \bar{S}_v^\nu\,,$$
$$\mathcal{M}_{21} = \sqrt{M_D M_F}\,\lambda_1\,\mathcal{A}^\mu + (iv\cdot\partial)^{-1}\,S_v^\mu \bar{T}_v\,,$$
$$\mathcal{M}_{22} = (\Delta_F^{\mu\nu})^{-1} + (iv\cdot\partial)^{-1}\,S_v^\mu \bar{S}_v^\nu\,,$$
$$\quad + \lambda_2\,\epsilon^{\mu\nu}{}_{\sigma\alpha}(\mathcal{A}^\sigma D^\alpha + h.c)\,, \qquad (30)$$

where $\Delta_D$ and $\Delta_F^{\mu\nu}$ denote the free propagators of the scalar and axial vector diquark respectively. The above determinant gives rise to an effective action, $\log\det[\cdots] = \mathrm{Tr}\log[\cdots]$. It can be expanded in terms of ordinary quark-diquark loops, resulting in an effective baryon lagrangian. This is done in the next section.

## 4 Loop expansion and effective lagrangian

In order to regularize the integrals appearing in the loop expansion of (29), we will use the proper-time method

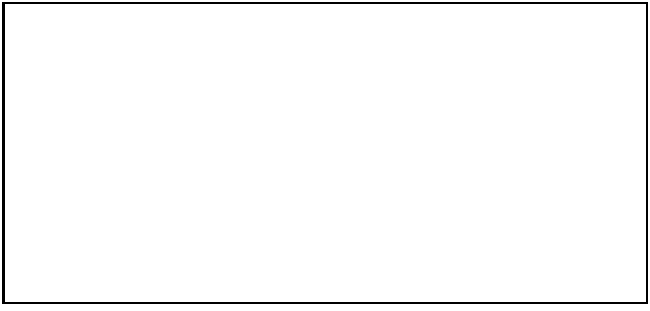

**Fig. 1.** Self–energy diagram for the baryon field $T_v$.

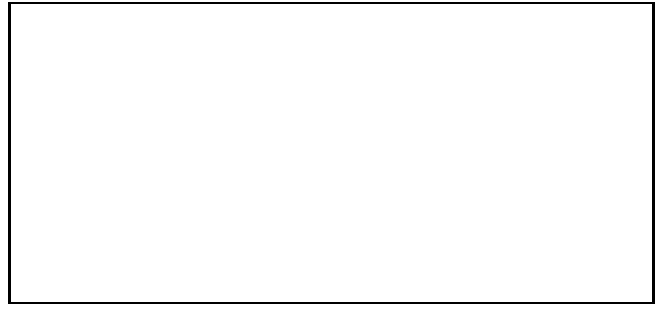

**Fig. 2.** Self–energy diagram for the baryon field $S_v^\mu$.

and rewrite the diquark propagator (i = D,F) after a Wick–rotation as

$$\frac{1}{k^2 + M_i^2} \to \int_{1/\Lambda^2}^{\infty} ds\, e^{-s(k^2+M_i^2)} . \tag{31}$$

All integrals will then be given in terms of the incomplete Gamma–function

$$\Gamma[\alpha, x] = \int_x^{\infty} dt\, e^{-t} t^{\alpha-1} . \tag{32}$$

### 1. Self energy for $T_v$

The self energy part for the spin 1/2 baryon field $T_v$ (15) following from (29) is represented by a Feynman diagram in Fig. 1 and formally reads

$$-i\,\mathrm{Tr}\left[\bar{T}_v (iv\cdot\partial)^{-1} T_v \Delta_D\right] . \tag{33}$$

Together with the quadratic term in (25) this contributes to the effective lagrangian

$$\mathrm{tr}\left[\bar{T}_v T_v\right]\left(-\frac{1}{\tilde{G}_1} + \Pi^T\right) \tag{34}$$

where $\Pi^T$ is given in the low–momentum expansion by

$$\begin{aligned}\Pi^T(v\cdot p) &= \frac{-3i}{16\pi^4}\int d^4k\, \frac{1}{v\cdot k + v\cdot p + i\epsilon}\cdot\frac{1}{k^2 - M_D^2 + i\epsilon}\\ &= I_3^D + 2I_2^D\, v\cdot p + O(v\cdot p)^2 ,\end{aligned} \tag{35}$$

with

$$I_2^i = \frac{3}{16\pi^2}\Gamma\left[0, M_i^2/\Lambda^2\right] , \tag{36}$$

$$I_3^i = \frac{3}{16\pi^2}\sqrt{\pi} M_i\, \Gamma\left[-1/2, M_i^2/\Lambda^2\right] . \tag{37}$$

From this we can read off the renormalization factor for the baryon field $T_v \to \sqrt{Z_T} T_v$ with $Z_T = (2I_2^D)^{-1}$, and the residual mass $\Delta M_T = M_T - m_Q$ which is independent of the heavy quark mass and now related to the coupling constant $\tilde{G}_1$ by

$$\Delta M_T = Z_T\left(\frac{1}{\tilde{G}_1} - I_3^D\right) . \tag{38}$$

### 2. Self energy for $S_v^\mu$

The term

$$-i\,\mathrm{Tr}\left[\bar{S}_v^\mu (iv\cdot\partial)^{-1} S_v^\nu \Delta_{F\,\mu\nu}\right] \tag{39}$$

arising from the loop expansion of (29) (see the Feynman diagram in Fig. 2) contributes to the self energy part for the baryon field $S_v^\mu$ (16) as

$$-\mathrm{tr}\left[\bar{S}_v^\mu S_v^\nu\right]\left(-\frac{1}{\tilde{G}_2} g_{\mu\nu} + \Pi^S_{\mu\nu}\right) . \tag{40}$$

Here, the loop contribution is given by

$$\begin{aligned}\Pi^S_{\mu\nu}(v\cdot p) &= g_{\mu\nu}\frac{-3i}{16\pi^4}\int d^4k\,\frac{1}{v\cdot k + v\cdot p + i\epsilon}\cdot\frac{1}{k^2 - M_F^2 + i\epsilon}\\ &\quad\times\left(1 - \frac{k_\perp^2}{3M_F^2}\right)\\ &= g_{\mu\nu}\left(I_3^F + \tilde{I}_3^F + (2I_2^F + 2\tilde{I}_2^F)\, v\cdot p\right) + O(v\cdot p)^2 ,\end{aligned} \tag{41}$$

where

$$\tilde{I}_2^i = \frac{3}{32\pi^2}\Gamma\left[-1, M_i^2/\Lambda^2\right] , \tag{42}$$

$$\tilde{I}_3^i = \frac{3}{32\pi^2}\sqrt{\pi} M_i\, \Gamma\left[-3/2, M_i^2/\Lambda^2\right] \tag{43}$$

and $k_\perp^\mu = k^\mu - (v\cdot k) v^\mu$. As before the proper normalization of the baryon field $S_v^\mu \to \sqrt{Z_S}\, S_v^\mu$ is performed by means of $Z_S = (2I_2^F + 2\tilde{I}_2^F)^{-1}$, and the residual mass $\Delta M_S = M_S - m_Q$ is now connected with the coupling $\tilde{G}_2$,

$$\Delta M_S = Z_S\left(\frac{1}{\tilde{G}_2} - I_3^F - \tilde{I}_3^F\right) . \tag{44}$$

### 3. Strong interactions with light mesons

The insertions of the axial currents into the diquark propagator generate the related axial couplings of the heavy baryons to the field $\mathcal{A}_\mu$. From the determinant (29) one obtains in the low–momentum expansion of the corresponding vertex diagrams in Fig. 3,

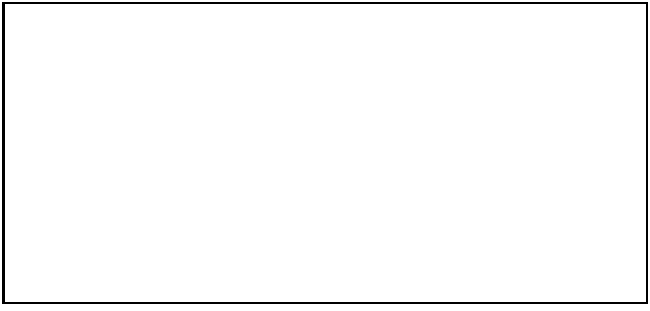

**Fig. 3.** Vertex diagram with insertion of the axial current $\mathcal{A}_\sigma$.

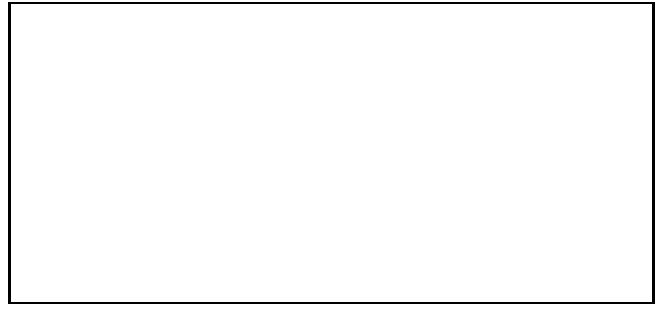

**Fig. 4.** Vertex diagram with insertion of a weak heavy quark current $\Gamma$.

$$-i\lambda_1 \sqrt{M_D M_F}$$
$$\times \, \mathrm{Tr}\left[(iv\cdot\partial)^{-1} S_v^\mu \bar{T}_v \Delta_{F\,\mu\nu} \mathcal{A}^\nu \Delta_D + h.c.\right], \quad (45)$$
$$-i\lambda_2$$
$$\times \, \mathrm{Tr}\left[(iv\cdot\partial)^{-1} S_v^\nu \bar{S}_v^\mu \Delta_{F\,\mu\mu'} \epsilon^{\mu'\sigma\alpha\beta}(2\mathcal{A}_\sigma\partial_\alpha)\Delta_{F\,\beta\nu}\right] (46)$$

This results in a renormalization of the axial coupling constants in the effective heavy baryon lagrangian

$$-\tilde{\lambda}_1 \,\mathrm{tr}\left[\bar{T}_v \mathcal{A}_\mu S_v^\mu + h.c.\right]$$
$$+\tilde{\lambda}_2 \, i\epsilon_{\mu\sigma\alpha\beta}\,\mathrm{tr}\left[\bar{S}_v^\mu \mathcal{A}^\sigma v^\alpha S_v^\beta\right], \quad (47)$$

with $\tilde{\lambda}_1 = \sqrt{Z_S Z_T}\, Z_1^{-1} \lambda_1$ and $\tilde{\lambda}_2 = Z_S Z_2^{-1} \lambda_2$. The vertex renormalization constants are calculated as

$$Z_1^{-1} = \sqrt{M_D M_F}\,\frac{3i}{16\pi^4}\int d^4k \frac{1}{v\cdot k + i\epsilon}$$
$$\times \frac{1}{k^2 - M_F^2}\frac{1}{k^2 - M_D^2}\left(1 - \frac{k_\perp^2}{3M_F^2}\right)$$
$$= \sqrt{M_D M_F}\,(I_4^{DF} + \tilde{I}_4^{DF}), \quad (48)$$

$$Z_2^{-1} = -\frac{3i}{16\pi^4}\int d^4k \frac{2}{(k^2 - M_F^2)^2} = 2I_2^F. \quad (49)$$

with

$$I_4^{DF} = \frac{3}{16\pi^2}\int_0^1 dx \sqrt{\frac{\pi}{xM_F^2 + (1-x)M_D^2}}$$
$$\times \Gamma\left[1/2, \frac{xM_F^2 + (1-x)M_D^2}{\Lambda^2}\right], \quad (50)$$

$$\tilde{I}_4^{DF} = \frac{3}{32\pi^2}\int_0^1 dx \sqrt{\pi}\,\frac{\sqrt{xM_F^2 + (1-x)M_D^2}}{M_F^2}$$
$$\times \Gamma\left[-1/2, \frac{xM_F^2 + (1-x)M_D^2}{\Lambda^2}\right]. \quad (51)$$

The coupling of the baryons to the light vector fields $\mathcal{V}_\mu$ is given by the covariant derivatives (19), (20). In other words the vector couplings do not suffer renormalization due to Ward–Identities for the $SU(3)_V$ symmetry.

The above results are summarized in the following effective baryon lagrangian

$$\mathcal{L}_{eff} = \mathrm{tr}\left[\bar{T}_v(iv\cdot\mathcal{D} - \Delta M_T)T_v\right]$$
$$-\mathrm{tr}\left[\bar{S}_{v\,\mu}(iv\cdot\mathcal{D} - \Delta M_S)S_v^\mu\right]$$
$$-\tilde{\lambda}_1 \,\mathrm{tr}\left[\bar{T}_v \mathcal{A}_\mu S_v^\mu + h.c.\right]$$
$$+\tilde{\lambda}_2 \, i\epsilon_{\mu\sigma\alpha\beta}\,\mathrm{tr}\left[\bar{S}_v^\mu \mathcal{A}^\sigma v^\alpha S_v^\beta\right]. \quad (52)$$

4. The Isgur–Wise form factors and comparison with Bjorken sum rule constraints

Due to the heavy spin symmetry, matrix elements of weak heavy quark currents $\bar{Q}_v \Gamma Q_{v'}$ in the heavy quark limit can be parametrized by means of a few unique form factors (Isgur–Wise functions) depending on the velocity transfer $(v\cdot v')$ [14]

$$A(v\cdot v')\,\bar{T}_v \Gamma T_{v'}, \quad (53)$$
$$G(v\cdot v')\,\bar{S}_v^\mu \Gamma T_{v'}\, v_\mu' + h.c., \quad (54)$$
$$\left\{B(v\cdot v')\,g_{\mu\nu} + C(v\cdot v')\,v_\mu' v_\nu\right\} \bar{S}_v^\mu \Gamma S_{v'}^\nu. \quad (55)$$

For $\Gamma = \gamma_\mu$ one obtains the matrix element of the flavour symmetry current and consequently the Isgur–Wise form factors $A(\omega)$ and $B(\omega)$ are normalized at zero recoil $A(1) = B(1) = 1$. The form factor $G(\omega)$ has to vanish identically, since there is no coupling to heavy quarks which mixes scalar and axial vector diquarks. The other form factors are predicted by simply inserting the weak current into the heavy quark line and calculating the one-loop Feynman diagrams (see Fig. 4). We obtain

$$A(v\cdot v') = Z_T\,\frac{3i}{16\pi^4}\int d^4k \frac{1}{v\cdot k + i\epsilon}\cdot\frac{1}{v'\cdot k + i\epsilon}$$
$$\times \frac{1}{k^2 - M_D^2 + i\epsilon}$$
$$= r(v\cdot v'), \quad (56)$$

$$r(\omega) = \frac{\ln(\omega + \sqrt{\omega^2 - 1})}{\sqrt{\omega^2 - 1}}, \quad (57)$$

and

$$B(v\cdot v')\,g_{\mu\nu} + C(v\cdot v')\,(v_\mu' v_\nu + v_\mu v_\nu')$$
$$= Z_S\,\frac{iN_C}{16\pi^4}\int d^4k \frac{1}{v\cdot k + i\epsilon}\cdot\frac{1}{v'\cdot k + i\epsilon}$$
$$\times \frac{1}{k^2 - M_F^2 + i\epsilon}\left(g_{\mu\nu} - \frac{k_\mu k_\nu}{M^2}\right), \quad (58)$$

$$B(\omega) = r(\omega) + Z_S\, 2\tilde{I}_2^F \frac{\omega - r(\omega)}{2 + \omega^2}, \quad (59)$$

$$C(\omega) = -Z_S\, 2\tilde{I}_2^F \frac{2 + \omega\, r(\omega)}{2 + \omega^2}. \quad (60)$$

The values at the non-recoil point $v = v'$ are $A(1) = B(1) = 1$ and $C(1) = -Z_S\, 2\tilde{I}_2^F$, and the three form factors in our model are related via the parameter-independent equation

$$B(\omega) = A(\omega) - \frac{\omega - r(\omega)}{2 + \omega\, r(\omega)} C(\omega) \,. \tag{61}$$

Phenomenologically it is important to know the slope of these form factors at zero recoil in order to correctly extrapolate the experimental data and to extract $|V_{cb}|$, one of the Standard Model parameters entering the CKM matrix. We obtain

$$A'(1) = -1/3 \,, \tag{62}$$

$$\begin{aligned} B'(1) &= -1/3 + 4/9\, Z_S\, 2 \tilde{I}_2^F \\ &= \{-0.14,\, -0.19,\, -0.22\} \,; \\ &\text{for } \frac{M_F^2}{\Lambda^2} = \left\{ \frac{1}{3},\, \frac{2}{3},\, 1 \right\} \,, \end{aligned} \tag{63}$$

$$C'(1) = B'(1) + 1/3 \,. \tag{64}$$

Although the parameter range of our model cannot be restricted sufficiently by experimental data it is worth comparing our results with the theoretical constraints imposed by Bjorken's sum rules [16]. Following the discussion in ref. [2], one obtains [4]

$$A(\omega) \le 1 \,, \qquad \left.\frac{dA}{d\omega}\right|_{\omega=1} \le 0 \,, \tag{65}$$

$$f(\omega) := \frac{2}{3}|B(\omega)|^2 + \frac{1}{3}\left|\omega\, B(\omega) + (\omega^2 - 1)\, C(\omega)\right|^2 \le 1 \,, \tag{66}$$

$$\left.\frac{dB}{d\omega}\right|_{\omega=1} \le -\frac{1}{3} - \frac{2}{3}\, C(1) \,. \tag{67}$$

with $\omega \ge 1$.

The function $A(\omega) = r(\omega)$ is obviously compatible with the bounds in eq. (65). Inserting our result for $B(\omega)$ and $C(\omega)$ into eq. (67) gives

$$\begin{aligned} \left.\frac{dB}{d\omega}\right|_{\omega=1} &= -\frac{1}{3} + \frac{4}{9} Z_S\, 2\, \tilde{I}_2^F \\ &< -\frac{1}{3} - \frac{2}{3}\, C(1) = -\frac{1}{3} + \frac{2}{3} Z_S\, 2\, \tilde{I}_2^F \end{aligned} \tag{68}$$

which is always true since $Z_S\, \tilde{I}_2^F$ is positive. The constraint of eq. (66) is also satisfied as can be seen in Fig. 5.

The result for the Isgur–Wise form factors $A(\omega)$, $B(\omega)$ and $C(\omega)$ are plotted in Figs. 6, 7, 8.

## 5 Summary and outlook

Focusing on the interplay between chiral and heavy quark symmetries, in the present paper we have investigated a phenomenological model for the dynamics within a heavy baryon. The two light quarks are approximated as effective diquarks and the model lagrangian contains their coupling to the heavy quark and to the axial chiral current as parameters.

Heavy baryon fields have been introduced by usual functional integration methods such that the diquark and heavy quark fields could be integrated out explicitly. The low–momentum expansion then yields an effective lagrangian of strongly interacting heavy baryons, where the values of the residual masses $\Delta M_{T,S}$ and the axial couplings $\tilde{\lambda}_{1,2}$ have been related to the input parameters of the model. Furthermore, we have calculated the three Isgur–Wise functions relevant for heavy baryon transitions and discussed their slope at the non-recoil point. In particular it turns out that the Isgur–Wise form factors of our model satisfy the constraints imposed by the Bjorken sum rules.

In conclusion, our model reproduces the phenomenological structures reflecting heavy quark and chiral symmetry. However, it would be desirable to start already with an effective model of quark flavor dynamics, like the Nambu–Jona–Lasinio model, which recently has been successfully extended to describe mesons containing a heavy quark [8]. The effective four-quark coupling of the NJL model includes an attractive interaction in the diquark channel, and the diquarks can be introduced as collective fields in close analogy to the meson case, determining their effective masses and couplings in terms of a few parameters.

For light baryons it has been shown that using the quark–diquark picture [11, 12, 13], one can reduce the three body problem to an effective two body problem, the NJL predictions then reproducing the mass spectrum of the lowest baryon states quite well [17, 18, 19]. Generalizing this approach to heavy quark systems would lead to an analogous equation of light and heavy quarks interacting with heavy and light diquarks via effective quark exchange, while the fundamental diquark properties like the masses and axial couplings $\lambda_1$, $\lambda_2$ will be fixed by the NJL model. Although the dynamics is more involved than in the present simple model, one expects similar symmetry properties. Investigations along these lines are in progress [20].

---

[4] Our notation translates into that of ref. [2] via $A(\omega) = f^{(0)}(\omega)$, $B(\omega) = g_1^{(0)}(\omega)$, $C(\omega) = -g_2^{(0)}(\omega)$.

This article was processed by the author using the LaTeX style file *cljour2* from Springer-Verlag.

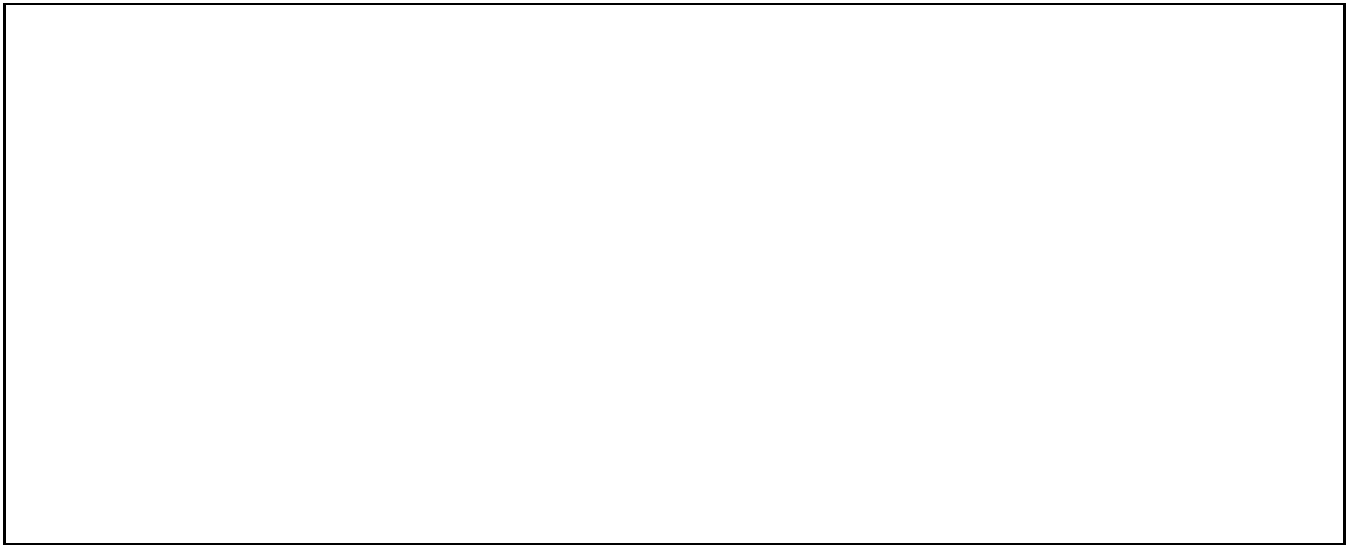

**Fig. 5.** The Bjorken bound $f(\omega) \leq 1$ plotted as a function of $\omega$ for three values $M_F^2/\Lambda^2 = 1/3,\ 2/3,\ 1$.

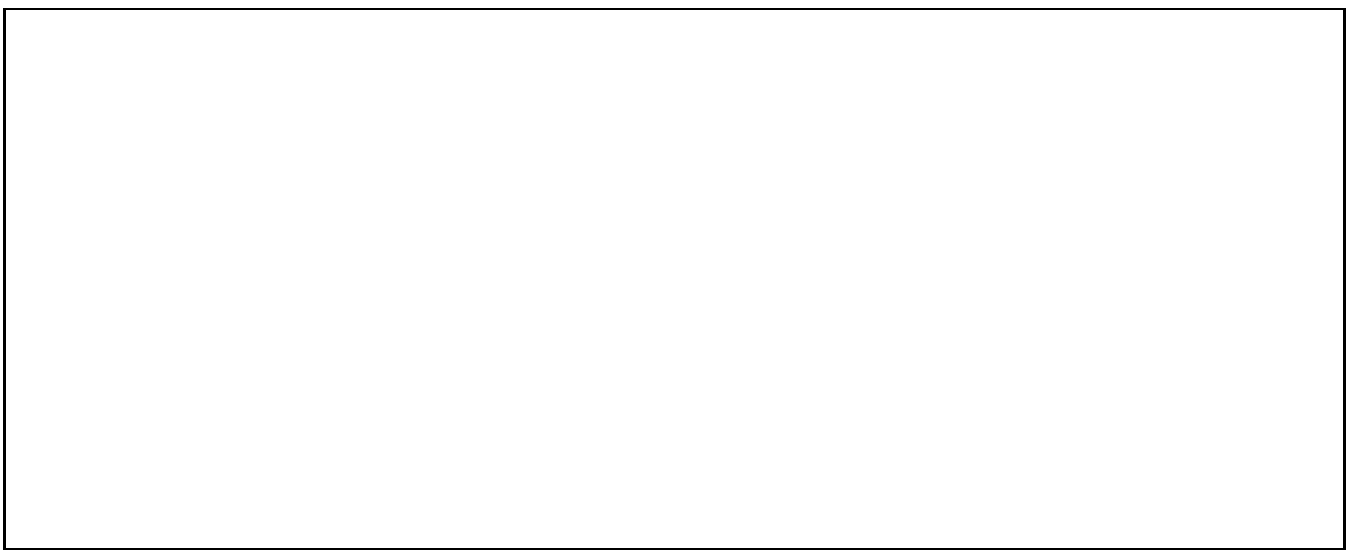

**Fig. 6.** The Isgur–Wise form factor $A(\omega)$ plotted as a function of $\omega$.

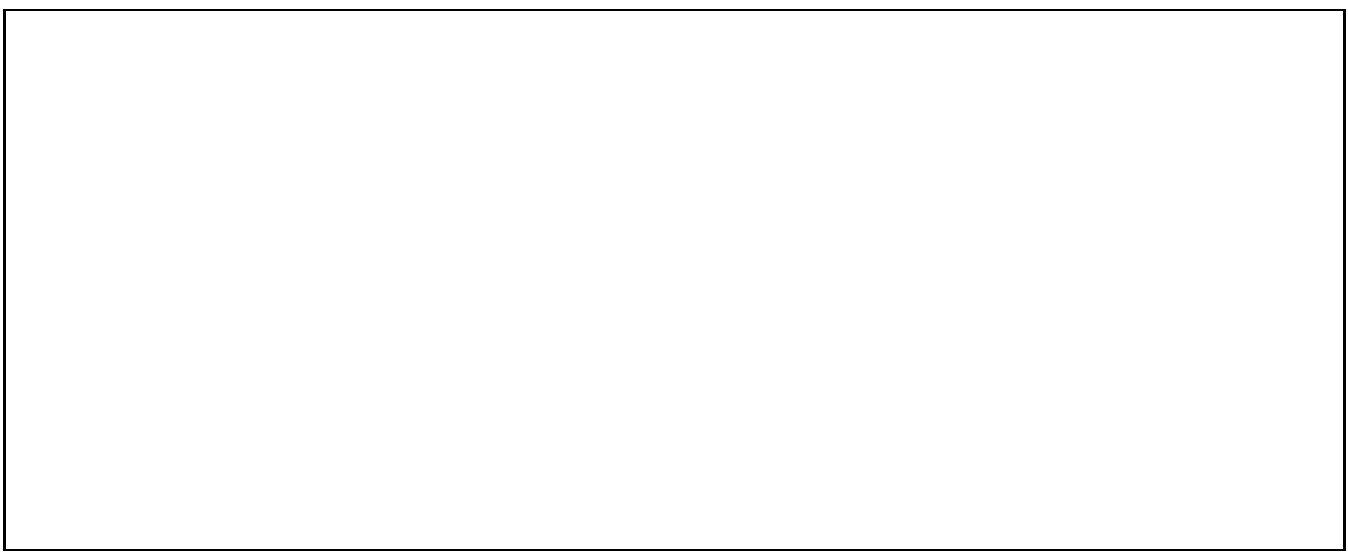

**Fig. 7.** The Isgur–Wise form factor $B(\omega)$ plotted as a function of $\omega$ for three values $M_F^2/\Lambda^2 = 1/3,\ 2/3,\ 1$.

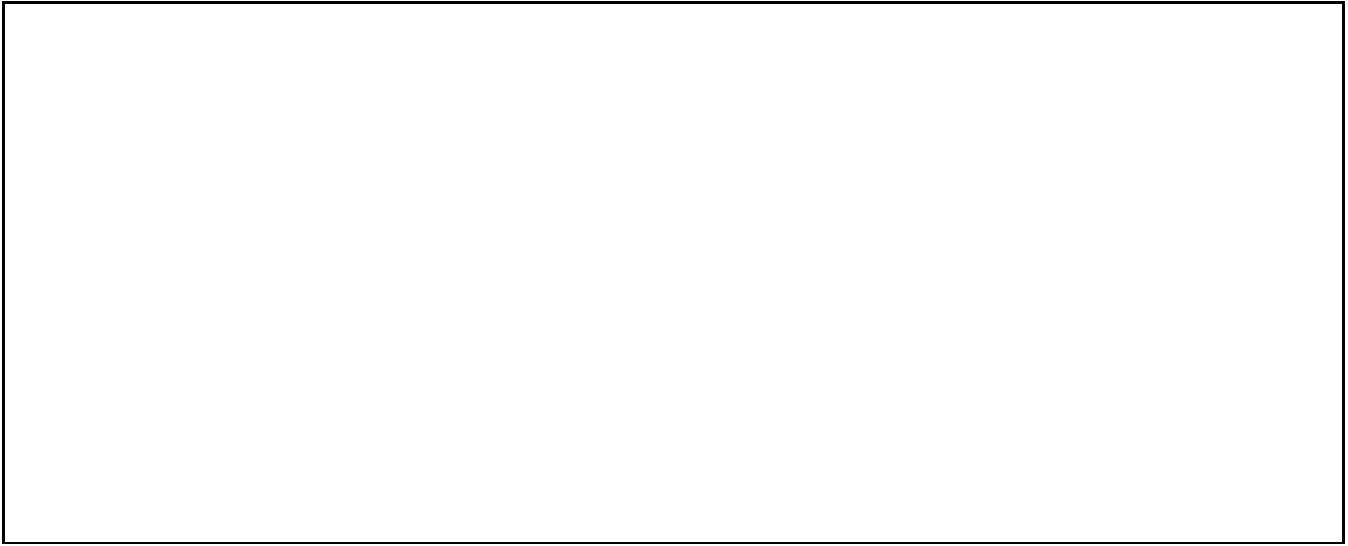

**Fig. 8.** The Isgur–Wise form factor $C(\omega)$ plotted as a function of $\omega$ for three values $M_F^2/\Lambda^2 = 1/3$, $2/3$, $1$.

# Figure Captions

**Figure 1:** Self–energy diagram for the baryon field $T_v$.

**Figure 2:** Self–energy diagram for the baryon field $S_v^\mu$.

**Figure 3:** Vertex diagram with insertion of the axial current $\mathcal{A}_\sigma$.

**Figure 4:** Vertex diagram with insertion of a weak heavy quark current $\Gamma$.

**Figure 5:** The Bjorken bound $f(\omega) \leq 1$ plotted as a function of $\omega$ for three values $M_F^2/\Lambda^2 = 1/3,\ 2/3,\ 1$.

**Figure 6:** The Isgur–Wise form factor $A(\omega)$ plotted as a function of $\omega$.

**Figure 7:** The Isgur–Wise form factor $B(\omega)$ plotted as a function of $\omega$ for three values $M_F^2/\Lambda^2 = 1/3,\ 2/3,\ 1$.

**Figure 8:** The Isgur–Wise form factor $C(\omega)$ plotted as a function of $\omega$ for three values $M_F^2/\Lambda^2 = 1/3,\ 2/3,\ 1$.

# Figures

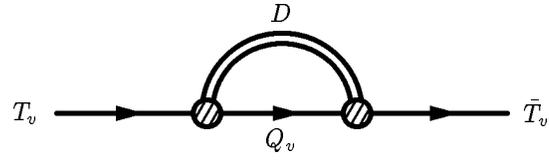

Figure 1: Self–energy diagram for the baryon field $T_v$.

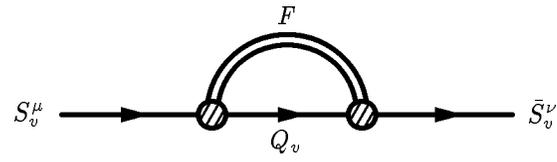

Figure 2: Self–energy diagram for the baryon field $S_v^\mu$.

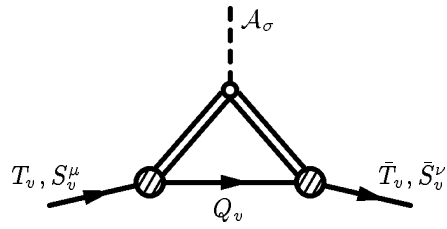

Figure 3: Vertex diagram with insertion of the axial current $\mathcal{A}_\sigma$.

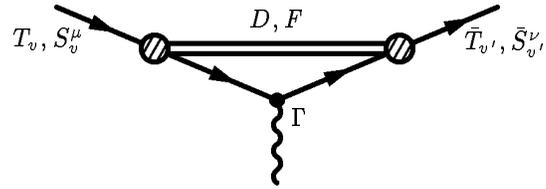

Figure 4: Vertex diagram with insertion of a weak heavy quark current $\Gamma$.

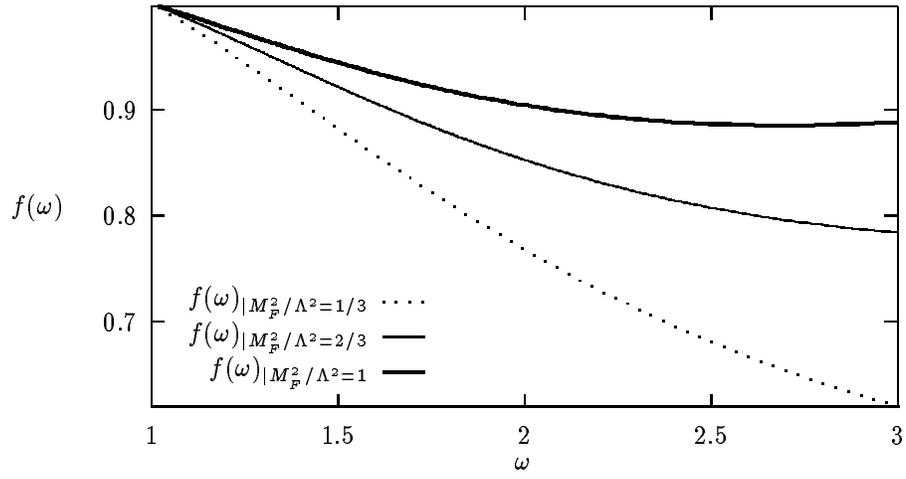

Figure 5: The Bjorken bound $f(\omega) \leq 1$ plotted as a function of $\omega$ for three values $M_F^2/\Lambda^2 = 1/3,\ 2/3,\ 1$.

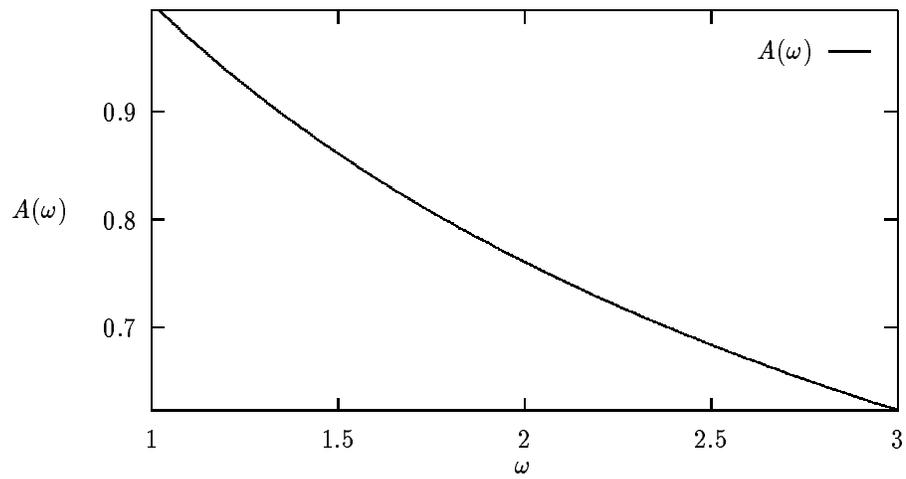

Figure 6: The Isgur–Wise form factor $A(\omega)$ plotted as a function of $\omega$.

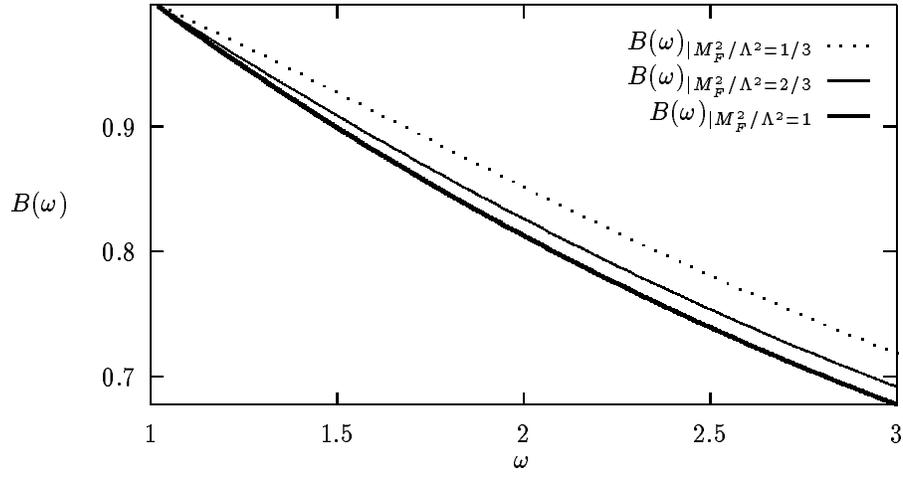

Figure 7: The Isgur–Wise form factor $B(\omega)$ plotted as a function of $\omega$ for three values $M_F^2/\Lambda^2 = 1/3$, $2/3$, $1$.

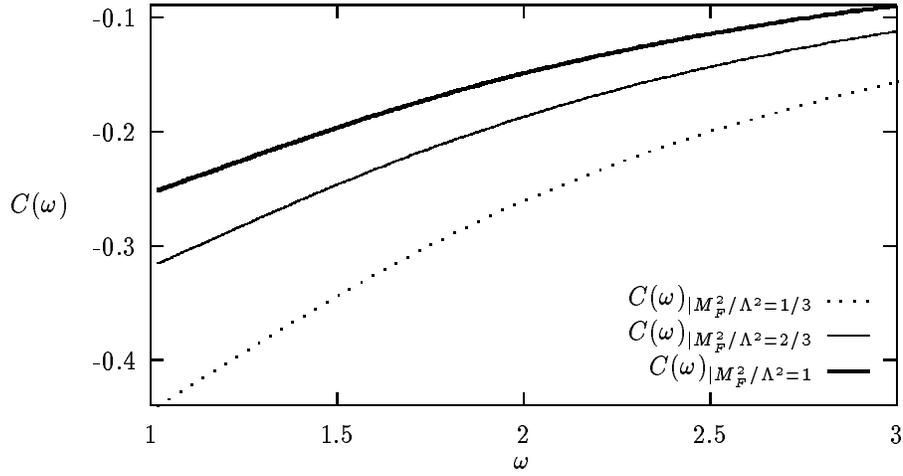

Figure 8: The Isgur–Wise form factor $C(\omega)$ plotted as a function of $\omega$ for three values $M_F^2/\Lambda^2 = 1/3$, $2/3$, $1$.